\documentclass[useAMS,usenatbib,fleqn,twocolumn]{aastex63}
\usepackage{amsmath}
\usepackage{multirow}
\usepackage{graphicx}
\usepackage{color,soul}
\usepackage{epsfig}
\usepackage{amssymb,amsmath}
\usepackage{aas_macros}
\begin{document}

\submitjournal{ApJL}
\shorttitle{RPS and AGN}
\title{A Link Between Ram Pressure Stripping and Active Galactic Nuclei}

\shortauthors{Ricarte et al.}
\correspondingauthor{Angelo Ricarte}
\email{angelo.ricarte@cfa.harvard.edu}

\author{Angelo Ricarte}
\affiliation{Center for Astrophysics | Harvard \& Smithsonian, 60 Garden Street, Cambridge, MA 02138, USA}

\author{Michael Tremmel}
\affiliation{Yale Center for Astronomy and Astrophysics, Physics Department, Yale University, P.O. Box 208120, New Haven, CT 06520, USA}

\author{Priyamvada Natarajan}
\affiliation{Department of Astronomy, Yale University, 52 Hillhouse Avenue, New Haven, CT 06511, USA}
\affiliation{Department of Physics, Yale University, P.O. Box 208121, New Haven, CT 06520, USA}

\author{Thomas Quinn}
\affiliation{Department of Astronomy, University of Washington, PO Box 351580, Seattle, WA 98195, USA}

\begin{abstract}
The dense environment of a galaxy cluster can radically transform the content of in-falling galaxies. Recent observations have found a significant population of active galactic nuclei (AGN) within ``jellyfish galaxies,'' galaxies with trailing tails of gas and stars that indicate significant ram pressure stripping. The relationship between AGN and ram pressure stripping is not well understood. In this letter, we investigate the connection between AGN activity and ram pressure in a fully cosmological setting for the first time using the {\sc RomulusC} simulation, one of the highest resolution simulations of a galaxy cluster to date. We find unambiguous morphological evidence for ram pressure stripping. For lower mass galaxies (with stellar masses $M_* \lesssim 10^{9.5} \ \mathrm{M}_{\odot}$) both star formation and black hole accretion are suppressed by ram pressure before they reach pericenter, whereas for more massive galaxies accretion onto the black hole is enhanced during pericentric passage. Our analysis also indicates that as long as the galaxy retains gas, AGN with higher Eddington ratios are more likely to be the found in galaxies experiencing higher ram pressure. We conclude that prior to quenching star formation, ram pressure triggers enhanced accretion onto the black hole, which then produces heating and outflows due to AGN feedback. AGN feedback may in turn serve to aid in the quenching of star formation in tandem with ram pressure. 
\end{abstract}


\section{Introduction}\label{sec:intro}

As clusters of galaxies assemble, they dynamically transform the physical properties of in-falling cluster members.  Galaxies falling into a cluster experience ram pressure from dense intra-cluster medium (ICM) gas that can potentially unbind their individual gas reservoirs \citep{Gunn&Gott1972}.  This process referred to as ram pressure stripping (RPS) can eventually remove a galaxy's entire gas supply, making it an important quenching pathway for satellite galaxies \citep{Vollmer+2001,Tonnesen+2007}.  Observationally, RPS results in disturbed galaxy morphologies and trailing tails of stripped gas \citep[e.g.,][]{vanGorkom2004,Kenney+2004,Cramer+2019}.  The most extreme examples have been dubbed ``jellyfish'' galaxies, due to the evocative morphologies of their star forming tails \citep{Ebeling+2014,Boselli+2016,Poggianti+2016}.  Prior to complete gas removal, moderate values of ram pressure have also been shown to increase the star formation rate, in galaxies both observed \citep{Crowl&Kenney2006,Merluzzi+2013,Vulcani+2018} and simulated \citep{Kronberger+2008,Tonnesen&Bryan2009,Kapferer+2009,Bekki2014}.  In this picture, the increased pressure initially helps compress the gas and triggers increased star-formation.  Over time, the interstellar medium (ISM) is fully stripped from the galaxy and star-formation ceases.  Recently, a very high incidence of AGN (5/7) has been observed in a sample of jellyfish galaxies \citep{Poggianti+2017}, and comprehensive follow-up of this sample has led to the identification of AGN-driven outflows \citep{Radovich+2019} and a compelling case for AGN feedback in action \citep{George+2019}.  It is plausible that the same gas compression that initially promotes star formation can also fuel active galactic nuclei (AGN), but this has not been demonstrated in hydrodynamical simulations to date.

In this work, we investigate the connection between ram pressure and AGN using a high-resolution hydro-dynamical cluster simulation for the first time.  {\sc RomulusC} is one of the highest resolution cosmological simulations of a galaxy cluster at present, containing innovative recipes for the seeding, accretion, and dynamics of super-massive black holes (SMBHs) \citep{Tremmel+2017a,Tremmel+2019}.  This cluster has a virial mass of $10^{14} \ \mathrm{M}_\odot$ and contains well-resolved satellite galaxies down to stellar masses of $10^8 \ \mathrm{M}_\odot$.  As we show, the high resolution of this simulation naturally produces ram pressure stripped tails as galaxies fall through the cluster.  We find that the end of a galaxy's star formation history is often punctuated by a brief increase in the SMBH accretion rate, which in turn heats the surrounding gas.  AGN triggered by the increased ram pressure produce observable outflows, which may be important for heating the densest gas in a galaxy deep down the potential well, and in turn facilitating the quenching process.  While both RPS and tidal stripping are expected to operate and cause dynamical transformations in galaxy clusters, tidal stripping effects are sub-dominant in the mass range of in-falling galaxies investigated here as noted in the Appendix. 

\section{Results}\label{sec:results}

\begin{figure*}
   \centering
   \includegraphics[width=\textwidth]{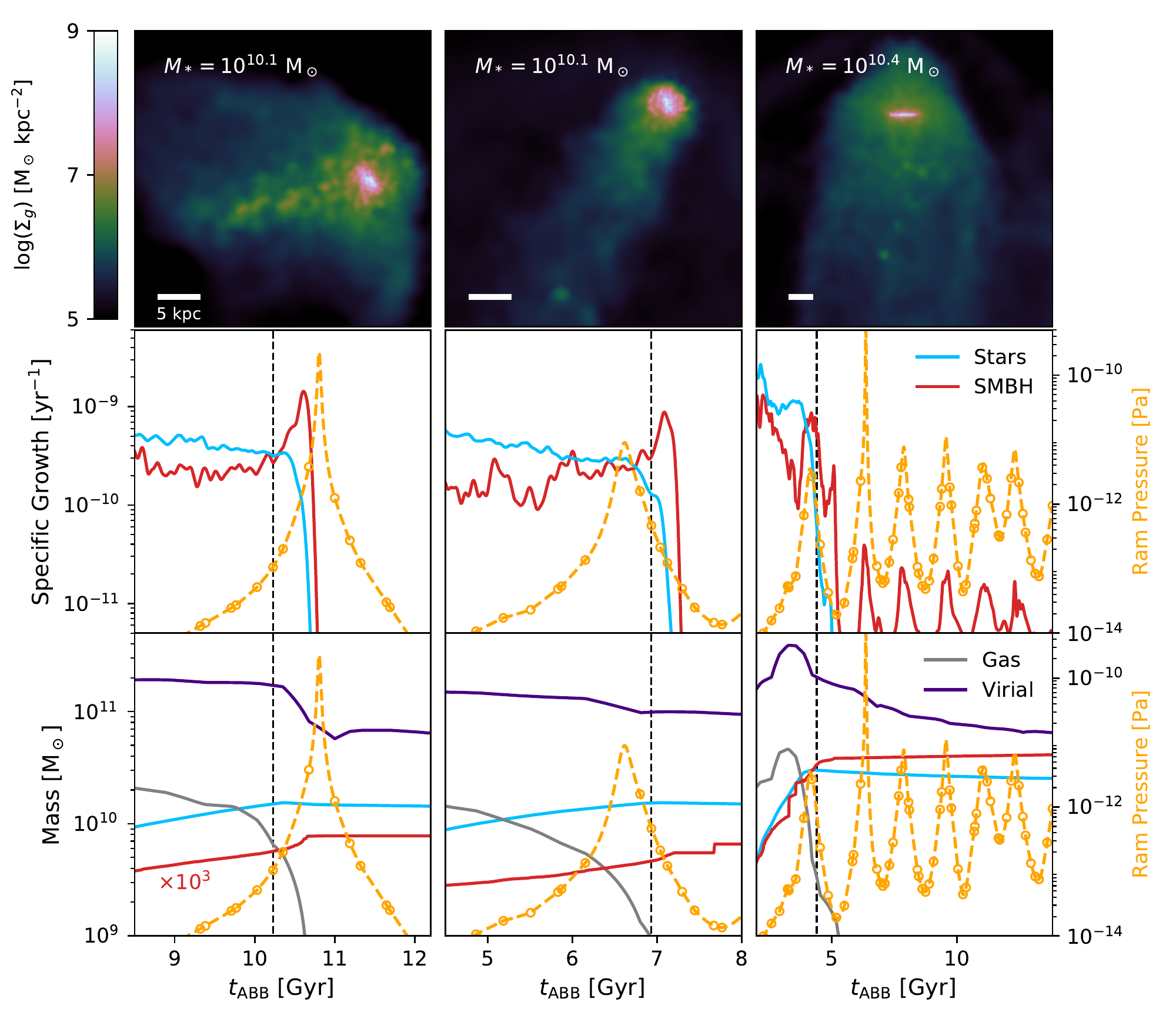}
   \caption{Three examples of enhanced AGN activity correlated with peaks in ram pressure during first pericenter passage in {\sc RomulusC}.  The top row shows a snapshot of the integrated surface mass density of the gas at the times indicated by the dashed vertical line in subsequent rows, revealing dramatic ram pressure stripped tails.  The middle row plots the specific BHAR and specific SFR for these galaxies over time. The BHAR peaks just prior to quenching, marking the end of RPS.  The bottom row plots the total stellar, SMBH ($\times 10^3$), virial, and bound gas mass.  Ram pressure is overplotted as orange dashed curves labeled on the right hand side.  Density profiles, positions, and velocities are saved at the times marked by open circles, which are interpolated onto the finer time resolution of the BHAR and SFR.  \label{fig:examples}}
  \end{figure*}

We find unambiguous morphological evidence of ram-pressure stripping along with AGN triggering among the members of {\sc RomulusC}.  Three illustrative examples are shown in Figure \ref{fig:examples}.  The top row displays gas column density maps during times marked with a dashed line in subsequent rows.  The middle row displays the specific BHAR and SFR, while the bottom row displays total stellar mass, SMBH mass ($\times 10^3$), gas mass, and virial mass.  For readability, specific BHAR and SFR are smoothed with a Gaussian kernel with a standard deviation of 30 Myr.  The orange dashed curves show the incident ram pressure inferred from Equation \ref{eqn:gunn&gott} \citep{Gunn&Gott1972}.  Since both ambient density and galaxy velocities increase with decreasing clustrocentric radius, a peak in ram pressure corresponds to a pericenter passage.  Open circles mark snapshot times at which densities and velocities are output; these curves are inferred by interpolating gas density profiles and position-velocity data over time.

In these three examples, the gas mass of in-falling cluster members is unbound within one orbit, impacted much more strongly by the cluster environment than either stellar or virial mass.  This confirms that RPS dominates over tidal stripping, which would remove dark matter and stars to a greater degree than gas.  The BHAR peaks during the final phases of RPS, while star formation in the galaxy quenches.\footnote{Note that a common definition of a quenched galaxy is a specific SFR below $10^{-11} \ \mathrm{yr}^{-1}$.}  Peaks in the specific BHAR occur near, but not exactly aligned with, peaks in ram pressure.  As we show in section \ref{sec:case_study}, AGN-driven winds help evacuate the final remnants of gas during the last phases of the stripping process.  In the third example, the most massive of the three, the galaxy is able to retain a very small amount of gas after the first orbit.  Each time the galaxy passes pericenter, it resumes low-level AGN activity (peak Eddington ratios of $\approx 10^{-2}$ to $10^{-3}$).  Similar behavior has been noted in a simulation of the formation of a compact elliptical galaxy (without a SMBH), resulting in central starbursts \citep{Du+2019}.  While this is the most dramatic example of repetitive SMBH fueling at pericenter passage, we notice similar behavior in three other cluster members, all of relatively high stellar mass.

\subsection{BHAR and SFR During Pericenter Passage}
\label{sec:agn_enhancement}

\begin{figure*}
   \centering
   \includegraphics[width=\textwidth]{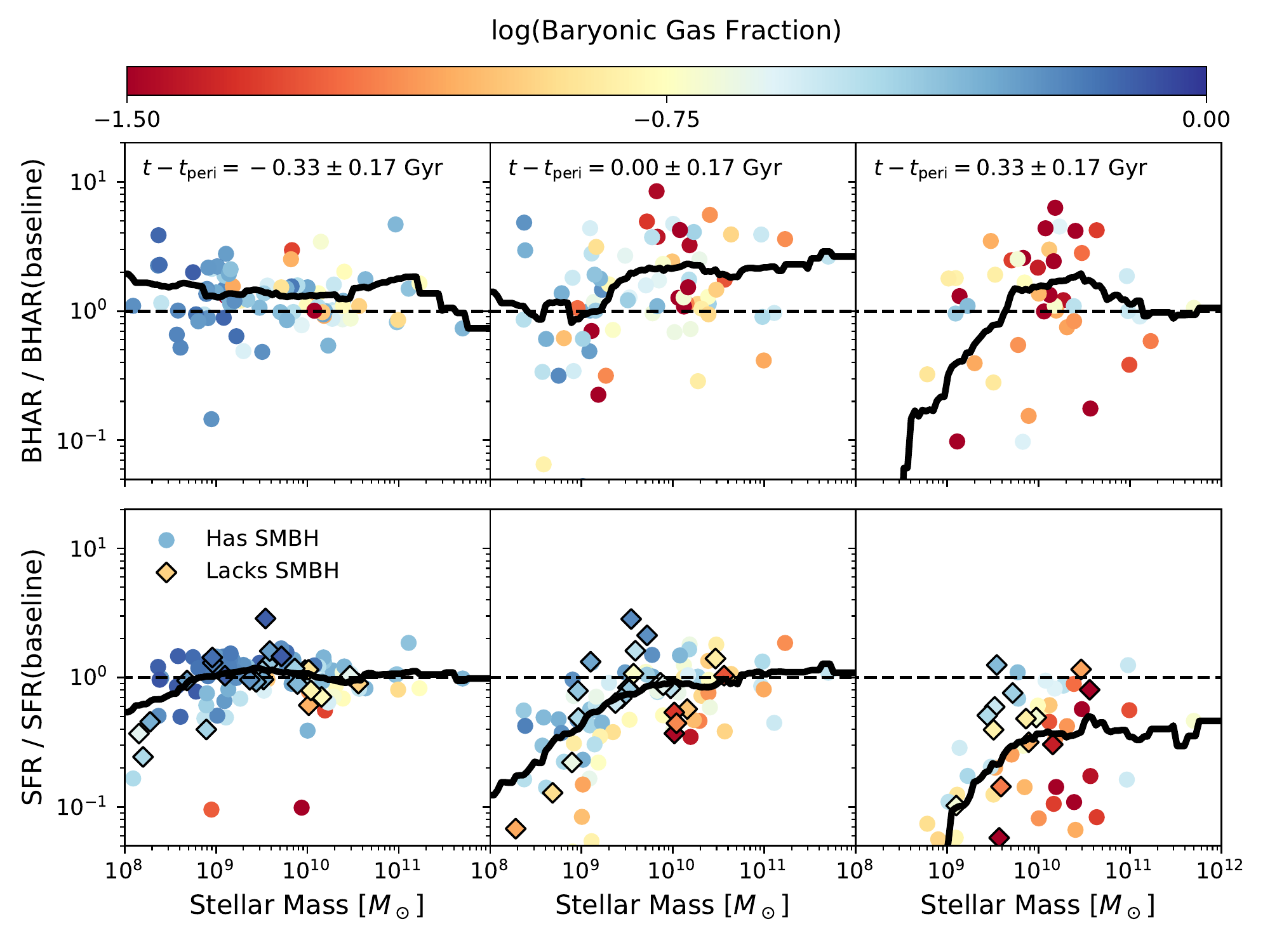}
   \caption{Evolution of BHAR and SFR during pericenter passage compared to baseline values 0.5-1.0 Gyr prior to pericenter passage.  The black line shows a running average with a window 1 dex in size in $M_*$.  During pericenter passage, for galaxies with stellar masses $>10^{9.5} \ \mathrm{M}_\odot$, the BHAR is enhanced on average by a factor of 2.2.  The SFR is suppressed more strongly in less massive galaxies. \label{fig:enhancement}}
\end{figure*}

In Figure \ref{fig:enhancement}, we examine the evolution of the BHAR and SFR around pericenter passage for the population of cluster members as a whole.  For each galaxy with a central SMBH, we first identify first pericenter passage by calculating the first time a galaxy experiences a ram pressure maximum with a peak value of at least $10^{-13}$ Pa.  Since galaxies sometimes have messy orbital interactions with other cluster members, this definition helps avoid uncertainties in defining pericenter based on minima in clustrocentric radius, for example.  We then average the BHAR and SFR for time periods before, during, and after pericenter passage, and take the ratio with respect to baseline values.  Baseline values are defined as the averages over 0.5 to 1.0 Gyr prior to pericenter passage, counting only times during which the galaxy has bound gas.  We remove one outlier:  a massive galaxy which is quenched prior to reaching its pericenter, but resumes forming stars at a modest rate during pericenter passage, resulting in a SFR enhancement of $7\times 10^4$.

Figure \ref{fig:enhancement} displays the evolution of BHAR and SFR throughout first pericenter passage, averaged over time periods noted on the top of each panel.  Points are color-coded according to the galaxy's baryonic gas fraction, defined as $M_g/(M_g+M_*)$ where $M_g$ and $M_*$ are the bound gas and stellar mass respectively.  The black curve represents a running average with a window size of 1 dex in stellar mass.  On average, the BHAR is enhanced by a factor of 1.6 during pericenter passage.  However, counting only galaxies above stellar masses of $10^{9.5} \ M_\odot$, the average enhancement rises to a factor of 2.2.  Below this mass, both the BHAR and SFR are already suppressed before pericenter.  Lower mass galaxies are more rapidly transformed by the RPS process than more massive galaxies with stronger gravitational potentials.  Interestingly, the two galaxies which exhibit the largest enhancements of the SFR during pericenter (seen clearly in the lower panel plots) both lack a central SMBH.  At later times, BHARs remain elevated in massive galaxies while the SFR is universally suppressed across a wide range of stellar masses.  This is consistent with the outside-in picture of RPS, which removes star-forming gas in the outskirts of galaxies earlier than it removes the central gas that fuels SMBHs.  

\subsection{BHAR and SFR During Early Ram Pressure Stripping}
\label{sec:correlations}

\begin{figure*}
   \centering
   \includegraphics[width=\textwidth]{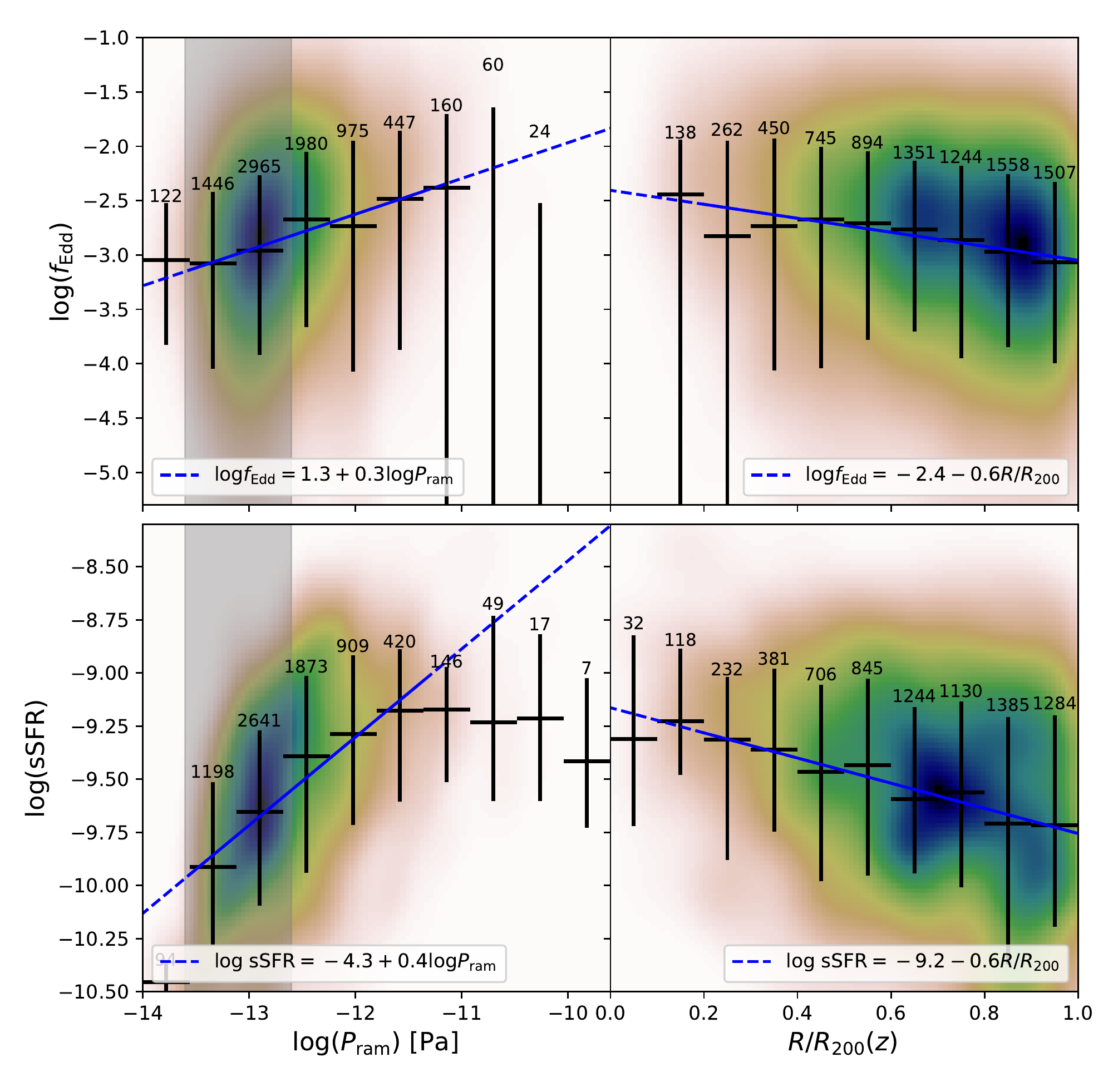}
   \caption{Eddington ratios ({\it top}) and specific star formation rates ({\it bottom}) versus ram pressure ({\it left}) and cluster-centric distance ({\it right}) for progenitors of $z=0$ cluster members.  We include only the initial phases of RPS, while the galaxy has at least 50\% of the gas with which it entered the cluster, intentionally excluding the later quenched stages.  Numbers above each bin indicate the number of data points in the bin.  The grey band is the region posited to trigger AGN by \citet{Marshall+2018}.  Clear correlations are found, consistent with ram pressure systematically elevating both sSFR and $f_{Edd}$.
   \label{fig:rp_fedd}}
\end{figure*}

Idealized wind tunnel simulations predict that the SFR may increase by factors of a few during the early phases of RPS as the gas is compressed \citep{Kronberger+2008,Tonnesen&Bryan2009,Kapferer+2009,Bekki2014}.  In addition, \citet{Marshall+2018} used a semi-analytic model to reproduce the spatial distributions of AGN in clusters and found the best agreement if AGN were triggered when $2.5 \times 10^{-14} \ \mathrm{Pa} < P_\mathrm{ram} < 2.5 \times 10^{-13} \ \mathrm{Pa}$ and $P_\mathrm{ram}/P_\mathrm{int} > 2$, where $P_\mathrm{int}$ is the galaxy's internal pressure support.  To test these results with {\sc RomulusC}, we attempt to find correlations between Eddington ratio, specific SFR, ram pressure, and cluster-centric radius by searching through all of the values spanned by the cluster members in {\sc RomulusC}.  First, we trace back the main progenitor branches of every cluster member with stellar masses above $10^8 \ M_\odot$ at $z=0$.  For this particular analysis, the BHAR is averaged over short 10 Myr intervals, and we treat each of these intervals as an independent point. In this way, we assume that these galaxies sample the physically allowed parameter space over time.  We only include time intervals during which the galaxy is within the cluster and retains at least 50\% of the gas mass with which it entered.  We impose this criterion to isolate the initial enhancement of BHAR with ram pressure, and to exclude the later stages when the lack of gas naturally leads to low $f_\mathrm{Edd}$ and sSFR.

In Figure \ref{fig:rp_fedd}, we plot Eddington ratio and specific star formation rate versus the host galaxy's incident ram pressure and cluster-centric distance.  The heat map represents the distribution of 10 Myr data points spanned by the galaxies in {\sc RomulusC} over time.  These distributions are then binned as a function of ram pressure, and the 16th, 50th, and 84th percentiles are marked with black error bars.  The numbers above these error bars indicate the number of points in the bin.  In these early phases, both $f_\mathrm{Edd}$ and sSFR clearly increase with increasing ram pressure.  These trends appear to plateau at the most extreme ram pressures.  However, we caution that these bins have comparatively far fewer data points, many of which come from the same individual galaxy histories, and interpolating galaxy positions may lead one to infer erroneously high ram pressures during pericenter.  Similar trends are seen in the right column with cluster-centric distance.  Higher values of $f_\mathrm{Edd}$ and sSFR are more likely at small cluster-centric distances, where the ram pressure is higher.  The grey bar shows the range of ram pressure proposed to trigger AGN by \citet{Marshall+2018}.  Rather than a threshold, our results appear more consistent with a smooth increase in $f_\mathrm{Edd}$ with incident ram pressure, extending to larger values of ram pressure. In blue, we plot the result of a linear regression to the bins containing at least 200 points.  We note that these distributions contain long, non-Gaussian tails towards low values of Eddington ratio.  Bearing this in mind, we fit only the binned $\pm 1\sigma$ values and treat the scatter as Gaussian rather than fitting the data points directly to avoid biasing results towards the most extreme outliers, those being galaxies with extremely low $f_\mathrm{Edd}$ or sSFR.  We obtain the following fits for ram pressure triggered AGN activity and star formation rate while a galaxy retains at least 50\% of its gas mass at infall:  

\begin{align*}
&\log_{10} f_\mathrm{Edd} = 1.3 + 0.3\log_{10}P_\mathrm{ram} \pm 0.9 \\ 
&\log_{10}\mathrm{sSFR} = -4.3 + 0.4\log_{10}P_\mathrm{ram} \pm 0.4 \\
&\log_{10}f_\mathrm{Edd} = -2.4 - 0.6(R/R_{200}) \pm 1.0 \\
&\log_{10}\mathrm{sSFR} = -9.2 - 0.6(R/R_{200}) \pm 0.5
\end{align*}

\noindent where the values following the $\pm$ sign are the average $1-\sigma$ scatter in dex amongst bins with at least 200 points, weighted by the number of points in the bin.

\section{A Case Study:  RPS With and Without a SMBH}
\label{sec:case_study}

Like most modern hydro-dynamical simulations of galaxy formation, {\sc RomulusC} includes AGN feedback, which is believed to be necessary to prevent gas from over-cooling in massive halos \citep[e.g.,][]{Croton+2006,Kormendy&Ho2013}.  It naturally acts on the gas deepest in the galaxy's potential well, which is the most difficult gas for RPS to remove.  By heating and driving winds that radially re-distribute this gas to the outer parts of the galaxy, AGN feedback can make this gas more susceptible to stripping. 

\begin{figure*}
   \centering
   \includegraphics[width=\textwidth]{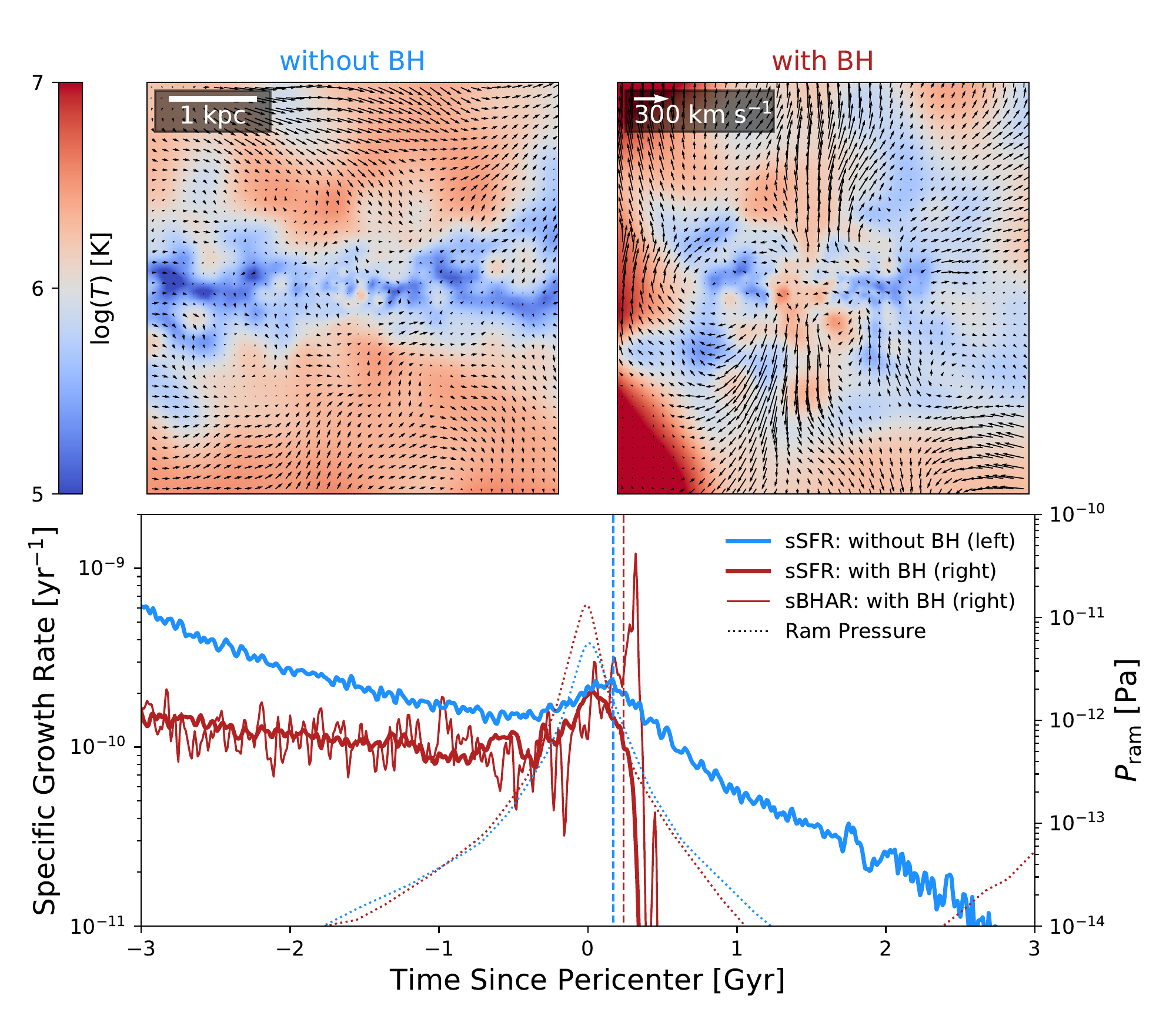}
   \caption{Case study of two similar galaxies with similar orbits, one with a central SMBH and one without.  The galaxy with a SMBH experiences an enhancement of AGN activity during pericenter and rapidly quenches.  The galaxy without a SMBH experiences a factor of $\approx 2$ enhancement in the SFR only, and quenches much more slowly over the course of 3 Gyr.  We compute temperature and velocity maps for each of these galaxies near pericenter passage, at the times marked by the dashed vertical lines. These maps reveal that AGN feedback in the galaxy with a central SMBH produces a hot outflow that helps rapidly quench the galaxy.
   \label{fig:case_study}}
\end{figure*}

{\sc RomulusC} produces a pair of similar galaxies that we study in more detail to better understand the relationship between ram pressure, quenching, and AGN.  These two galaxies each have stellar masses of $10^{10.5} \ M_\odot$ and experience similar ram pressure over time, but one has a central SMBH and the other does not.  The galaxy lacking a central SMBH only has wandering SMBHs, the most central of which is 17 kpc away at $z=0$.  While rare, this situation arises in {\sc RomulusC} because seeding is based on local rather than halo properties, and SMBHs are not forced to stay at the centers of their halos \citep{Tremmel+2018}. We compare the two galaxies in this case study in Figure \ref{fig:case_study}.  We plot specific SFR as thick solid lines, and the specific BHAR of the galaxy with a SMBH with a thin line.  Thin dotted lines depict the ram pressure, which is remarkably similar for these two galaxies.  At the times shown by the thin dashed lines, we plot temperature and velocity maps in the panels above.  As these galaxies pass through pericenter, both galaxies exhibit a mild enhancement of the specific SFR.  The galaxy lacking a central SMBH slowly quenches over the course of 3 Gyr.  In contrast, the galaxy containing a SMBH experiences a burst of AGN activity shortly after pericenter passage.  In the top row, we plot temperature and velocity maps taken after pericenter within 1 kpc wide-slabs that are oriented perpendicular to the disk.  These maps reveal that the AGN drives a hot outflow with speeds of hundreds of $\mathrm{km} \; \mathrm{s}^{-1}$, while no outflow develops in the galaxy without a central SMBH.  In this case, AGN feedback heats the gas, drives a wind, disturbs the disk morphology, and may help quench the galaxy on a much shorter time-scale.  This strongly resembles the jellyfish galaxy JO201, which has the same stellar mass, contains a central hole of molecular gas, and exhibits an outflow of 261 km s$^{-1}$ \citep{George+2019,Radovich+2019}.

\section{Conclusions}\label{sec:conclusions}

We have analyzed the connection between ram pressure and AGN activity in the state-of-the-art {\sc RomulusC} cosmological simulation of a $10^{14} \ \mathrm{M}_\odot$ cluster, one of the highest resolution clusters ever simulated.  We find that during peaks of ram pressure, the BHAR tends to increase by a factor of $\sim 2$, even while the SFR is in decline.  As long as a galaxy has gas, both the Eddington ratio and specific star formation rate in galaxies increase along with the incident ram pressure.  In one case study, RPS and AGN appear to work in synergy to quench star formation.  Ram pressure compresses the gas and helps fuel the AGN, which imparts energy to its surroundings to help facilitate the stripping process.  This suggests that AGN feedback may play a role in rapidly quenching galaxies during the final phases of RPS. 

Controlled, idealized wind-tunnel experiments of RPS containing AGN would be useful in better understanding these processes.  In cosmological simulations like {\sc RomulusC}, it is difficult to disentangle simultaneous processes, and many ``sub-grid'' approximations are required to make the problem computationally feasible.  Some important approximations and confounding factors are listed below:
\begin{itemize}
    \item The ISM:  {\sc RomulusC} lacks the resolution to fully resolve the patchy nature of the ISM.  High-resolution RPS simulations indicate that stripping proceeds more efficiently with a multi-phase ISM, since wind can move through the holes in the disk \citep{Quilis+2000,Tonnesen&Bryan2009}. 
    \item Galactic Magnetic Fields:  RPS simulations including a magnetic field have been shown to create more flared disks.  Oblique shocks that develop between the disk and the ICM promote gas inflows to the centers of galaxies, which may help fuel AGN more efficiently \citep{Ramos-Martinez+2018}. since {\sc RomulusC} lacks magnetic fields, we may under-estimate the AGN accretion events triggered in this way.
    \item SMBH Fueling and Feedback:  At present, neither SMBH fueling or feedback are fully understood, and there exist a variety of plausible models implemented in galaxy-scale simulations.  It would be useful to determine how robust our results are to variations of these models.  In particular, mechanical ``radio-mode'' feedback, which is not implemented in {\sc RomulusC}, is thought to be be much more efficient in removing gas, especially at low-Eddington ratios \citep[e.g.,][]{Choi+2015}.  This could potentially increase the effects of AGN feedback that we see at the end of the RPS process.
    \item Pre-processing:  Most galaxies that fall into {\sc RomulusC} are satellites in smaller substructures.  For example, it is known that {\sc RomulusC} experiences a group merger at $z\approx 0.2$ that disrupts the cluster's cool core \citep{Chadayammuri+2020}.  The total ram pressure experienced by a galaxy may therefore be underestimated since we do not take other substructures into account.
\end{itemize}

This work explains the high incidence of AGN among jellyfish galaxies \citep{Poggianti+2017} and reproduces the observed signatures attributed to AGN feedback \citep{George+2019,Radovich+2019}.  Note that {\sc RomulusC} is a $10^{14} \ \mathrm{M}_\odot$ cluster, while observed clusters with jellyfish galaxies can be up to an order-of-magnitude more massive.  We predict that in clusters of higher mass or redshift, the effects described in this work would likely occur at larger radius, due to higher ambient gas densities.  We speculate that this may be relevant for the observed high-incidence of AGN in the outskirts of high-redshift, massive clusters \citep{Koulouridis&Bartalucci2019}.  This work motivates more detailed and controlled studies to better understand the RPS-AGN connection using alternative techniques and sub-grid models for AGN and galaxy physics.

\section*{acknowledgements}  We thank Daisuke Nagai and Mila Chadayammuri for illuminating discussions about {\sc RomulusC} and other galaxy clusters.   AR acknowledges support by NASA Headquarters under the NASA Earth and Space Science Fellowship Program - Grant 80NSSC17K0459, and the National Science Foundation under Grant No. OISE 1743747.  AR also thanks the Black Hole Initiative for its computational resources and collaborative atmosphere.  PN acknowledges support from the National Science Foundation under TCAN 1332858.  MT gratefully acknowledges the support of the Yale Center for Astronomy and Astrophysics Postdoctoral Fellowship. 

This research is part of the Blue Waters sustained petascale computing project, which is supported by the National Science Foundation (awards OCI-0725070 and ACI1238993) and the state of Illinois. Blue Waters is a joint effort of the University of Illinois at Urbana-Champaign and its National Center for Supercomputing Applications. This work is also part of a Petascale Computing Resource Allocations allocation support by the National Science Foundation (award number OAC-1613674). This work also used the Extreme Science and Engineering Discovery Environment (XSEDE), which is supported by National Science Foundation grant number ACI-1548562. Resources supporting this work were also provided by the NASA High-End Computing (HEC) Program through the NASA Advanced Supercomputing (NAS) Division at Ames Research Center.  Analysis was conducted on the NASA Pleiades computer and facilities supported by the Yale Center for Research Computing.

\appendix
\section{Numerical Methods}\label{sec:methodology}

We analyze the {\sc RomulusC} simulation, a state-of-the-art zoom-in simulation of a $10^{14} \ \mathrm{M}_\odot$ galaxy cluster that includes AGN feedback \citep{Tremmel+2015,Tremmel+2017a}.  In previous work, we established that SMBHs and their galaxies co-evolve in the {\sc Romulus} simulations, irrespective of stellar mass, redshift, or intergalactic environment \citep{Ricarte+2019}.  \citet{Sharma+2019} investigate the scatter about these trends for dwarf galaxies, finding that SMBH growth is affected by halo assembly history and that feedback from AGN can affect the evolution of galaxies even at low masses. \citet{Tremmel+2019} find that the cluster environment results in significantly higher quenching rates on all mass scales.  Here, we briefly summarize the important and relevant gas and SMBH physics, while more details can be found in \citet{Tremmel+2017a,Tremmel+2019}.

The {\sc RomulusC} simulations are performed using the Tree + Smoothed Particle Hydrodynamics (SPH) code {\sc ChaNGa} \citep{Menon+2015}.  Gravity is softened below 350 pc using a spline kernel (equivalent to 250 pc Plummer softening), hydro-dynamics are resolved to 70 pc, and gas particles have masses of $2.12 \times 10^5 \ \rm{M}_\odot$.  While this is sufficient resolution to produce resolved stripped tails characteristic of ``jellyfish'' galaxies, it is not high enough to fully resolve the patchy structure of the ISM.  Magnetic fields are not included in these simulations.

SMBHs are seeded based on local gas properties, rather than simply imposing a halo mass threshold as is often adopted \citep[e.g.,][]{Springel+2005,Vogelsberger+2013,Schaye+2015}.  A gas particle designated to form a star instead forms a SMBH if it has low metallicity, high density, and a temperature of $\approx 10^4 \ \mathrm{K}$, motivated by models of direct collapse black hole seeding \citep{Haiman&Loeb2001,Oh&Haiman2002,Lodato&Natarajan2006,Begelman+2006}.  SMBHs grow from surrounding gas particles based on Bondi-Hoyle-Lyttleton \citep{Bondi1952} accretion, modified to account for angular momentum support. There is only one mode of AGN feedback: 2\% of the radiative energy released during the accretion process (assuming a radiative efficiency of 10\%) is dumped thermally into surrounding gas particles.  Finally, unlike many other cosmological simulations, SMBHs are not forced by hand to stay in gravitational potential minima. Their dynamics are self-consistently followed by correcting for unresolved dynamical friction due to gravitational softening \citep{Tremmel+2015}.

\section{Post-processing}

The Amiga Halo Finder \citep{Knollmann&Knebe2009} is used to identify substructure---halos, subhalos, and the galaxies at their centers. At $z=0$, in {\sc RomulusC} there exist 227 cluster members with stellar masses of at least $10^8 \ M_\odot$ in halos with at least $10^4$ dark matter particles.  We trace back the histories of every cluster member of {\sc RomulusC} using the tools available in the {\sc TANGOS} database structure \citep{Pontzen&Tremmel2018}, which uses particle tracking to identify halo merger trees. The black hole accretion rate (BHAR) and star formation rate (SFR) are each saved with a resolution of 10 Myr, while most other properties are only saved at each larger snapshot, of which there are 72 spaced between $z=12.88$ and $z=0$.  Importantly, these less frequently output quantities include each galaxy's coordinates and gas particle data.  These are interpolated onto the same time resolution as the BHAR and SFR.  To estimate the magnitude of ram pressure, we use the formula of \citet{Gunn&Gott1972}:
\begin{eqnarray}
P_\mathrm{ram} \approx \rho_{\rm ICM} (R) v^2 \label{eqn:gunn&gott}
\end{eqnarray}
\noindent where $\rho_{\rm ICM} (R)$ is the ambient gas mass density of the ICM, $R$ is the distance of the in-falling galaxy with respect to the cluster center, and $v$ the velocity of the in-falling galaxy relative to cluster center. To avoid ambiguities in the computation of these quantities locally, we compute $\rho_{\rm ICM}$ based on spherically averaged density profiles of the cluster gas (which excludes gas in satellite galaxies), and compute $v$ relative to that of the center of the dark matter halo of the cluster, which is computed using a shrinking spheres method.  We note that tidal stripping is expected to become more important than RPS only in galaxies which are at least $\approx 1/8$ the mass of the cluster \citep{McCarthy+2008}.  For {\sc RomulusC} with a cluster mass of $10^{14}\, \mathrm{M}_\odot$ and galaxies with the total masses in the range $10^{9-12.5}\,\mathrm{M}_\odot$ (with stellar masses in the range of $10^{8-11}\,\mathrm{M}_\odot$) considered here, we expect RPS effects to dominate over tidal stripping.

\bibliography{ms}

\end{document}